\documentclass[reprint,amsmath,amssymb,aps,prl,superscriptaddress]{revtex4-2}

\usepackage{hyperref}
\usepackage{xcolor}
\usepackage{graphicx}
\usepackage[alsoload=hep]{siunitx}
\usepackage{float}
\usepackage{physics}
\usepackage{makecell}
\usepackage{outlines}

\begin{document}

%\title{The Measurement-induced Transition in Circuits with Long-range Interactions}
\title{The Measurement-induced Transition in  Long-range Interacting Quantum Circuits}

\author{Maxwell Block}
\affiliation{Department of Physics, University of California, Berkeley, California 94720, USA}

\author{Yimu Bao}
\affiliation{Department of Physics, University of California, Berkeley, California 94720, USA}

\author{Soonwon Choi}
\affiliation{Department of Physics, University of California, Berkeley, California 94720, USA}
\affiliation{Center for Theoretical Physics, Massachusetts Institute of Technology, Cambridge, MA 02139, USA}

\author{Ehud Altman}
\affiliation{Department of Physics, University of California, Berkeley, California 94720, USA}
\affiliation{Materials Sciences Division, Lawrence Berkeley National Laboratory, Berkeley, California 94720, USA}

\author{Norman Y. Yao}
\affiliation{Department of Physics, University of California, Berkeley, California 94720, USA}
\affiliation{Materials Sciences Division, Lawrence Berkeley National Laboratory, Berkeley, California 94720, USA}

\date{\today}

%Despite being forbidden in equilibrium, spontaneous breaking of time translation symmetry can occur in periodically driven, Floquet systems with discrete time-translation symmetry. The period of the resulting discrete time crystal is quantized to an integer multiple of the drive period, arising from a combination of collective synchronization and many body localization. Here, we consider a simple model for a one dimensional discrete time crystal which explicitly reveals the rigidity of the emergent oscillations as the drive is varied. We numerically map out its phase diagram and compute the properties of the dynamical phase transition where the time crystal melts into a trivial Floquet insulator. Moreover, we demonstrate that the model can be realized with current experimental technologies and propose a blueprint based upon a one dimensional chain of trapped ions. Using experimental parameters (featuring long-range interactions), we identify the phase boundaries of the ion-time-crystal and propose a measurable signature of the symmetry breaking phase transition.The universality of this measurement-induced transition. Systems of strongly interacting dipoles offer an attractive platform to study many-body localized phases, owing to their long coherence times and strong interactions. We explore conditions under which such localized phases persist in the presence of power-law interactions and supplement our analytic treatment with numerical evidence of localized states in one dimension.

\begin{abstract}
The competition between scrambling unitary evolution and projective measurements leads to  a phase transition in the dynamics of quantum entanglement.
Here, we demonstrate that the nature of this transition is fundamentally altered by the presence of long-range, power-law interactions.
For sufficiently weak power-laws, the measurement-induced transition is described by conformal field theory, analogous to short-range-interacting hybrid circuits. 
However, beyond a critical power-law, we demonstrate that long-range interactions give rise to a continuum of non-conformal universality classes, with continuously varying critical exponents. 
We numerically determine the phase diagram for a one-dimensional, long-range-interacting hybrid circuit model as a function of  the power-law exponent and the measurement rate.  
Finally, by using an analytic mapping to a long-range quantum Ising model, we provide a theoretical understanding for the critical power-law.
\end{abstract}

\maketitle

%The emergence of platforms capable of supporting many-body entanglement have opened the door to explorations of many-body quantum dynamics in strongly interacting systems.

%The emergence of controllable quantum simulators supporting many-body entanglement is enabling, and motivating, new studies of previously inaccessible strongly-interacting dynamical systems.

\begin{figure}
    \centering
    \includegraphics[width=3.0in]{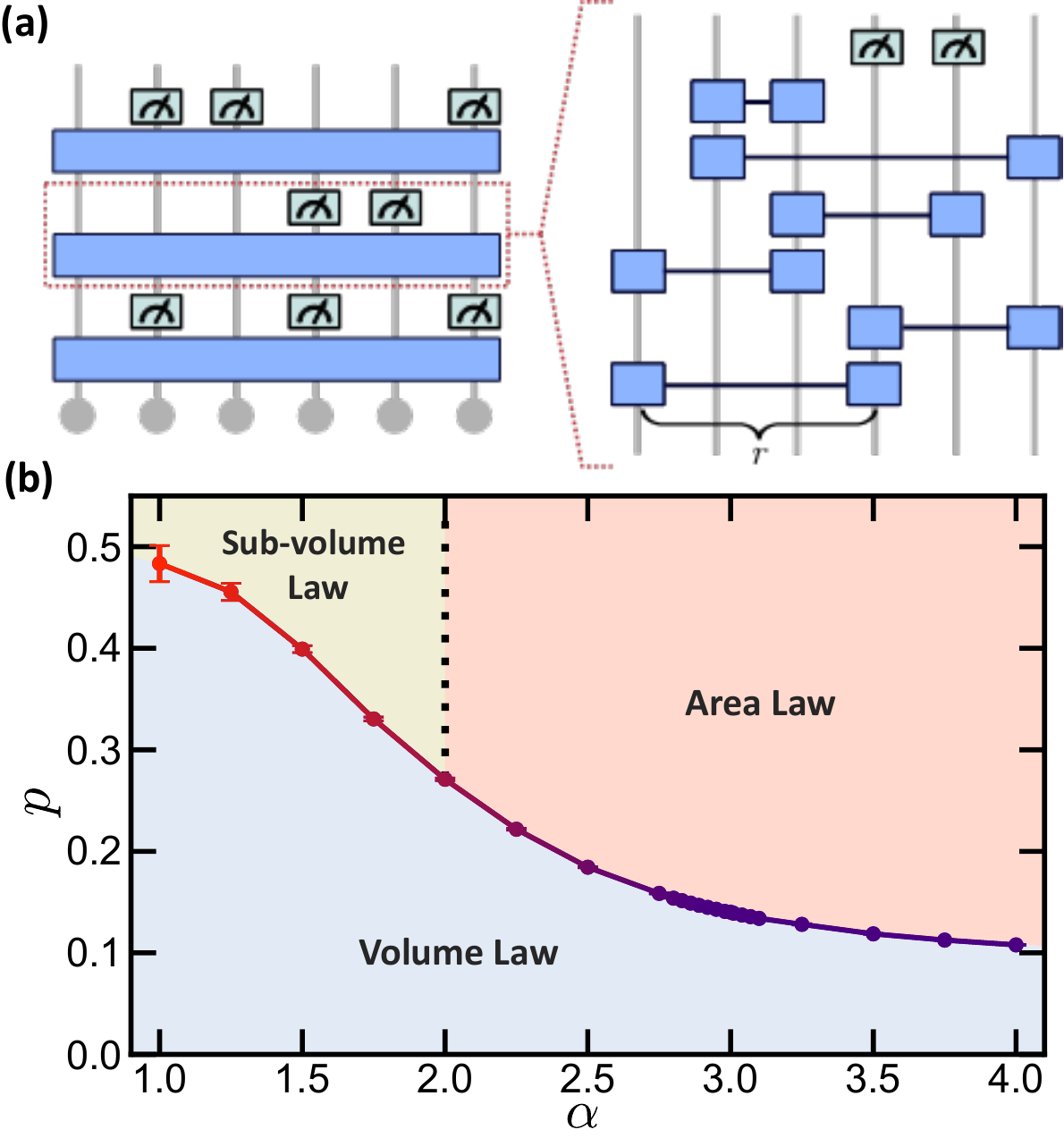}
    \caption{\small (a) Schematic of our long-range interacting hybrid circuit, which consists of interleaved layers of unitary evolution and randomly placed projective measurements. Two-qubit gates separated by distance $r$ occur with a probability $P(r)\sim 1/r^\alpha$.
    (b) Phase diagram as a function of the measurement rate, $p$ and the power-law exponent, $\alpha$. 
    For $\alpha \gtrsim 3$, the measurement-induced phase transition is described by conformal field theory (purple), while for $\alpha \lesssim 3$, the universality changes continuously (purple-red  gradient). 
    For $\alpha<2$, the area law phase is supplanted by a sub-volume law phase, where half-chain entanglement entropy ($S_{L/2}$) scales as $L^{2-\alpha}$.}
\label{fig:1}
\end{figure}

Programmable simulators---capable of supporting many-body entanglement---have opened the door to a new family of quantum dynamical questions~\cite{preskill_quantum_2018, arute_quantum_2019, gross_quantum_2017, bluvstein_controlling_2021, semeghini_probing_2021, satzinger_realizing_2021}. 
A unifying theme behind these queries is the competition between many-body entangling interactions and various types of entanglement-suppressing dynamics.
For example, many-body localization arises when interactions are pitted against strong  disorder \cite{abanin_colloquium_2019,nandkishore_many-body_2015,vosk_theory_2015, choi_exploring_2016}.
Similarly, the dissipative preparation of entangled states requires a delicate balance between unitary and incoherent evolution \cite{poyatos_quantum_1996, kraus_preparation_2008, shankar_autonomously_2013, carusotto_photonic_2020, shao_dissipative_2017}.
Recently, a tremendous amount of excitement has focused on a new paradigm for such competition, namely,  ``hybrid'' quantum circuits composed of scrambling dynamics interspersed with projective measurements (Fig.~\ref{fig:1})~\cite{li_quantum_2018,li_measurement-driven_2019,skinner_measurement-induced_2019,chan2019unitary,choi_quantum_2020}.

Naively, such evolution appears similar to the perhaps more familiar case of open-system dynamics, where an environment is viewed as constantly (at least weakly) measuring the system.
But there is a crucial difference: in open-system dynamics, the results of the environment's measurements are unknown, and only the average over outcomes determines the system's evolution~\cite{carmichael_open_1993, gardiner_quantum_2004, clerk_introduction_2010}.
In hybrid quantum circuits, however, the projective measurement results are recorded, so the dynamics resolve individual quantum trajectories~\cite{li_quantum_2018,skinner_measurement-induced_2019}.
This distinction has a profound consequence on the long-time dynamics.
%\cite{skinner_measurement-induced_2019, li_quantum_2018,choi_quantum_2020,gullans_dynamical_2020,gullans_scalable_2019,zabalo_critical_2020, li_statistical_2021, gullans_localization_2019, fan_self-organized_2020, ippoliti_entanglement_2021,piroli_random_2020,vasseur_entanglement_2019, lavasani_measurement-induced_2021, cao_entanglement_2019,gullans_entanglement_2019, gullans_localization_2019, szyniszewski_entanglement_2019, tang_measurement-induced_2020} 

Most fundamentally, instead of approaching a steady-state density matrix, the system perpetually fluctuates in Hilbert space, building up many-body entanglement that is, possibly, later eradicated by a few well-placed measurements~\cite{gullans_dynamical_2020,gullans_scalable_2019,zabalo_critical_2020, li_statistical_2021, gullans_localization_2019, fan_self-organized_2020}.
This constant ebb and flow of entanglement gives rise to a novel dynamical phase transition: at low measurement rates, the dynamics generate extensive entanglement, while at high measurement rates, only few-body entangled clusters emerge \cite{li_quantum_2018, li_measurement-driven_2019, skinner_measurement-induced_2019}.
To date, this measurement-induced transition has been explored in two limits: hybrid quantum circuits with local interactions \cite{vasseur_entanglement_2019, lavasani_measurement-induced_2021, cao_entanglement_2019,gullans_entanglement_2019, gullans_localization_2019, szyniszewski_entanglement_2019, tang_measurement-induced_2020} and all-to-all interacting circuits where powerful analytic techniques can be applied~\cite{nahum_measurement_2020, vijay_measurement-driven_2020,piroli_random_2020}. 
Understanding the nature of the measurement-induced transition in generic, long-range-interacting systems (i.e.~with power-laws $\sim 1/r^\alpha$) remains an essential open question that finds motivation from two complementary angles. 

First, such long-range interactions are known to have profound effects on the universality, and indeed, even  the existence, of many phase transitions~\cite{thouless_long-range_1969, mermin_absence_1966, hohenberg_existence_1967, maghrebi_continuous_2017, shastry_exact_1988, haldane_exact_1988}; in addition, long-range interactions can parametrically alter the form of Lieb-Robinson bounds and scrambling light-cones \cite{tran_hierarchy_2020, chen_quantum_2019, chen_finite_2019, else_improved_2020}.
%1
%Consequently, it is natural, even expected, for long-range interactions to dramatically affect the nature of the measurement-induced phase transition.
 %
Second, many of the most promising experimental platforms for investigating the measurement-induced transition, including Rydberg tweezer arrays, polar molecules, trapped ions and solid-state magnetic dipoles, inherently feature long-range interactions \cite{saffman_quantum_2010, demille_quantum_2002, monroe_programmable_2021, davis_probing_2021, hall_detection_2016}.
%
%Therefore, understanding the consequences of such long-range interactions is also of practical importance for realizing the measurement-induced phase transition in experiments. 

\begin{figure}
    \centering
    \includegraphics[scale=0.5]{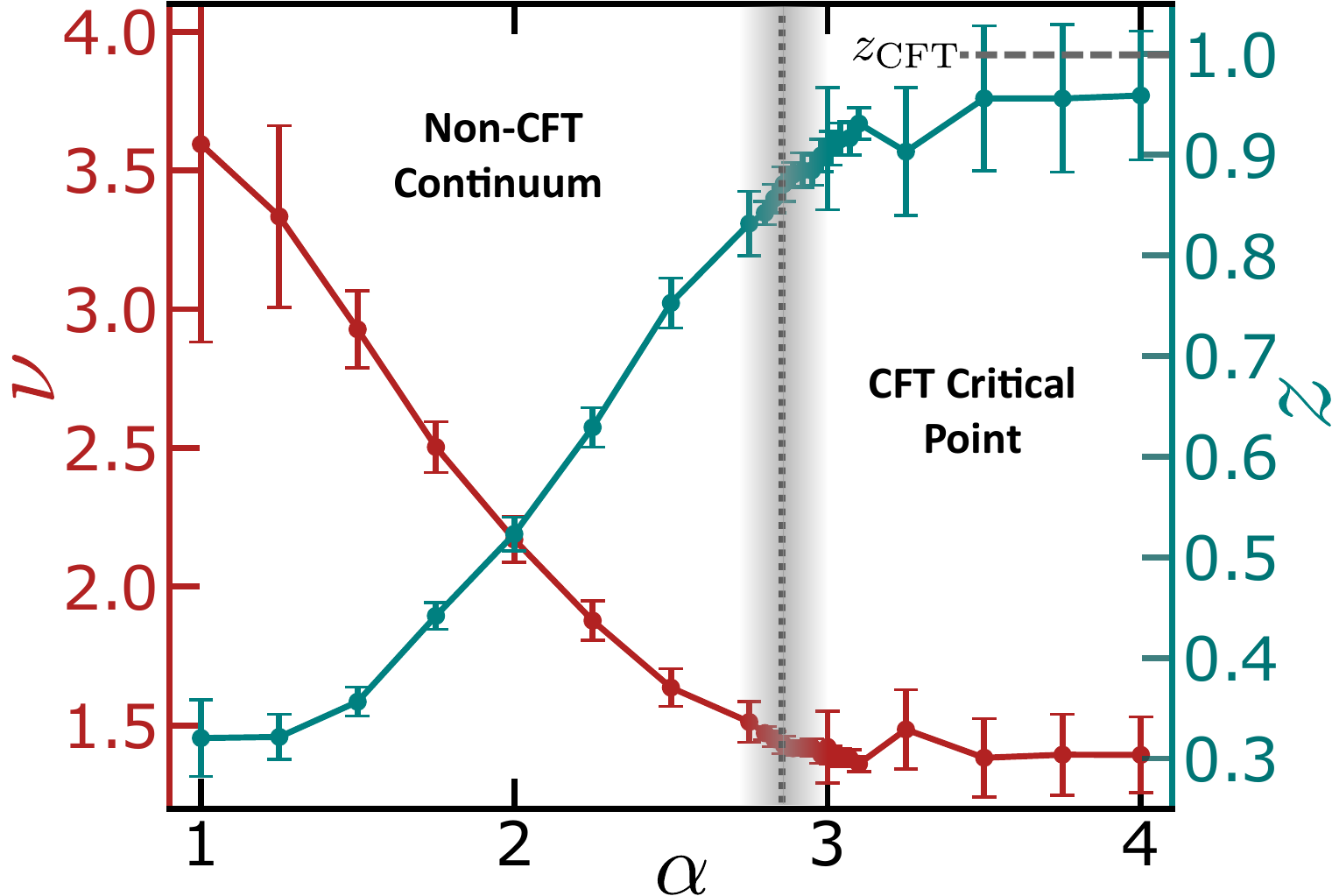}
    \caption{\small The correlation length critical exponent, $\nu$ (red), and the dynamical critical exponent, $z$ (teal), extracted from a finite-size scaling analysis of the purification time. 
    For $\alpha \gtrsim 3$, one finds $z \approx 1$ corresponding to a CFT. 
    For $\alpha \lesssim 3$, both critical exponents vary continuously, indicating a continuum of non-CFT universality classes in this regime. 
    The dotted line (and shaded grey region) is consistent with a critical power-law, $\alpha_\textrm{c} = 3 - \eta$, where $\eta \sim 0.2$ is the anomalous dimension.  
    }
\label{fig:2}
\end{figure}

In this Letter, we demonstrate that the interplay between long-range interactions and projective measurements  leads to fundamentally new universality classes for the measurement-induced transition. 
Our main results are three-fold.
First, we find that  for $\alpha \gtrsim 3$ the universality class  is consistent with previous studies of short-range models; however, for $\alpha \lesssim 3$, the phase transition is no longer described by conformal field theory (CFT) and exhibits continuously varying critical exponents (Fig.~\ref{fig:2}). 
Second, we determine the phase diagram associated with the transition as a function of the measurement rate, $p$, and the power-law exponent $\alpha$ [Fig.~\ref{fig:1}(b)].
For $\alpha>2$, the transition occurs between volume- and area-law phases, while for $\alpha<2$ \cite{skinner_measurement-induced_2019}, the area-law phase is supplanted by a ``sub-volume" law phase~\cite{supp}. 
Finally, we develop an exact correspondence between hybrid quantum circuits with long-range interactions and a quantum Ising model with long-range interactions. 
This correspondence allows us to understand the measurement-induced transition in terms of the ground-state properties of a quantum spin chain~\cite{sak_recursion_1973, dutta_phase_2001, defenu_criticality_2017}; perhaps most intriguingly, it provides an analytic explanation for the dramatic change in universality at $\alpha \approx 3$---this is precisely when long-range interactions become a relevant perturbation.
%
% We also explain the change of universality at $\alpha=3$ in terms of recent results on Lieb-Robinson bounds of long-range Hamiltonians \cite{chen_finite_2019, tran_hierarchy_2020}. 

% The unitary evolution consists of $L/2$ two qubit gates, acting on qubits separated by distance $r$ where $r$ is distributed according to $r \sim 1/r^\alpha$. 
\begin{figure*}
    \centering
    \includegraphics[scale=0.53]{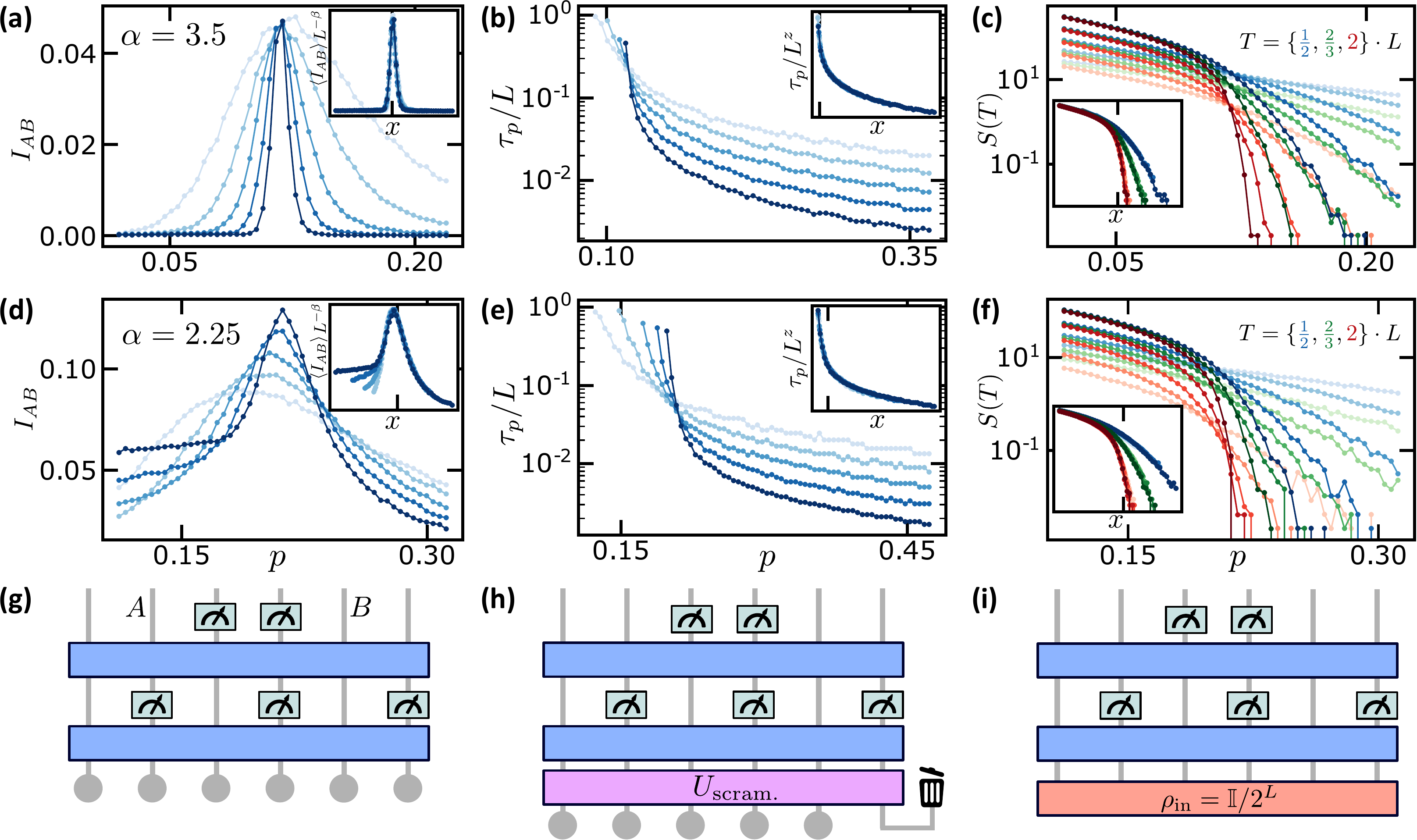}
    \caption{\small (a-c) The antipodal mutual information $I_{AB}$, purification time $\tau_p$, and global entropy dynamics $S(t)$ as a function of the measurement rate $p$, for power-law $\alpha=3.5$.   
    Insets depict the corresponding finite-size scaling collapse with $x=(p-p_c)L^{1/\nu}$ (the single tick mark on the x-axis denotes $x=0$).
    Different system sizes ($L=[32, 64, 128, 256, 512]$) are indicated via increasing opacity.
    (d-f) Depict analogous plots for $\alpha=2.25$. Both the peak-heights of $I_{AB}$ [d] and the crossing points of $\tau_p/L$ [e] exhibit marked $L$-dependence. This 
  immediately indicates that the measurement-induced transition is no longer conformal.
   (c,f) Colors indicate different time-slices of $S(t)$ (according to the legend). 
   Finite-size collapses (insets) are obtained by rescaling  $t=cL \rightarrow cL^z$, with $c = \{1/2, 2/3, 2\}$ depending on the time-slice. 
   (g) Circuit schematic for $I_{AB}$. The system is initialized in a product state and the mutual information is measured between antipodal regions $A$ and $B$.
   (h) Circuit schematic for $\tau_p$. The system is initialized in a product state with a single maximally mixed qubit. To avoid early-time finite-size effects, we apply a global scrambling Clifford, $U_\mathrm{s}$, before evolving with our hybrid circuit. (i) Circuit schematic for $S(t)$. The system is initialized in a maximally mixed state and slowly purifies under hybrid dynamics.}
\label{fig:3}
\end{figure*}
% Note that although $\expval{I_{AB}}$ may be collapsed with a single parameter $\nu^*$, it does not accurately probe $\nu$ for $\alpha<3$ [see footnote].
% The system is initialized with one maximally mixed qubit in an otherwise pure product state, and is subsequently scrambled by a random Clifford $U$. The system is then evolved with our hybrid model until purification, and $\tau_p$ is determined as the median purification time.
% In order to avoid complications arising from ``lucky" measurements at finite size, i.e. the case where a maximally mixed qubit is measured early on in the dynamics, we ensure the bit of entropy is delocalized by creating from tracing out a bell-pair entangled with some scrambled state [Fig.~\ref{fig:1}(b)].
%
% To be precise, we define $\tau_p$ as the median time it takes for a such a single delocalized qubit of entropy to purify \footnote{We note the purification time $\tau_p$ is closely related to the average entropy of a reference qubit, an order parameter introduced in \cite{gullans_scalable_2019}.}.

\emph{Long-range hybrid quantum circuits.---}Consider a one-dimensional system of $L$ qubits with periodic boundary conditions.
Our hybrid quantum circuits consist of long-range gates interspersed with projective measurements [Fig.~\ref{fig:1}(a)] \cite{1_foot_clifford}.

%\footnote{We consider Clifford gates to make the simulation of large system systems practically feasible. The dynamics of random Clifford circuits are qualitatively to similar to those of Haar-random circuits \cite{zabalo_critical_2020}.}. 
%
More precisely, a single time step of the \emph{scrambling} portion of the evolution consists of $L$ random two-qubit Clifford gates acting on qubits separated by $r$ sites, with a probability distribution, $P(r) \sim 1 / r^\alpha$; each scrambling time-step is then followed by $pL$ projective measurements (randomly distributed)~\cite{supp}.

We have carefully chosen our scrambling dynamics to be qualitatively similar to those generated by long-range interacting Hamiltonians.
Indeed, the light cone (as measured via an out-of-time-order correlator) for our random circuit model with power-law  $\alpha$ is expected to match the corresponding light-cone generated by \emph{chaotic Hamiltonian} dynamics with power-law  $\alpha/2$~\cite{2_foot_op_spread}. 
%\footnote{To summarize the argument given in \cite{zhou_operator_2020}: operator spreading in chaotic dynamics is determined by a local coupling strength, $J$ and a local decoherence time, $\tau$,  with operators spreading occurring at a rate $J^2 / \tau$. If $J ~ 1/r^\alpha$, then the random circuit model with similar operator-spreading properties should have power-law exponent $2 \alpha$}.
%
To this end, our analysis also provides insights into the measurement-induced transition when the dynamics are driven by a long-range-interacting Hamiltonian (provided one maps $\alpha \rightarrow \alpha/2$).

\emph{Diagnostics.---}We characterize the dynamics of our long-range hybrid quantum circuits using four diagnostics:  1) the half-chain entanglement entropy ($S_{L/2}$); 2) the anti-podal mutual information ($I_\mathrm{AB}$)~\cite{li_measurement-driven_2019,skinner_measurement-induced_2019}; 3) the global purification dynamics ($S(t)$)~\cite{gullans_dynamical_2020}; and 4) the single-qubit purification time ($\tau_p$)~\cite{gullans_scalable_2019}. 
All observables are defined as average quantities over many circuit realizations, and $S_{L/2}$ and $I_{AB}$ are \emph{steady-state} quantities that are also averaged over late times. %, which we denote by angled brackets, $\expval{\cdot}$.

The half-chain entanglement entropy, $S_{L/2}$, is an intuitive diagnostic of the transition in the case of short-range interactions: at low measurement rates, the system evolves to an extensively entangled state and $S_{L/2} \sim L$ (volume law), while at high measurement rates the system remains in a product state and $S_{L/2} \sim O(1)$ (area law).
Due to sub-leading corrections to its critical scaling form, $S_{L/2}$ is challenging to work with quantitatively \cite{li_measurement-driven_2019}.
It turns out to be more straightforward to analyze $I_\mathrm{AB}$, defined as the mutual information between two small anti-podal regions [Fig.~\ref{fig:3}(g)].
Crucially, $I_\mathrm{AB} \approx 0$ in both the product \emph{and} extensively-entangled phases (where the system is unentangled or self-thermalizing respectively), and only peaks in the critical region, making it simple to use for finite-size scaling~\cite{li_measurement-driven_2019}.

Both $S_{L/2}$ and $I_\mathrm{AB}$ require a notion of geometric locality to be well-defined, which breaks down as $\alpha \rightarrow 0$ \cite{vijay_measurement-driven_2020, nahum_measurement_2020}. % are defined as \emph{steady-state} quantities, and are therefore not-sensitive to correlations in time, or more precisely, the dynamical exponent $z$.
Thus, in order to gain a complete understanding of the dynamics, we  also consider $\tau_p$, the median time it takes for measurements to purify a single qubit  [Fig.~\ref{fig:3}(h)] \cite{gullans_scalable_2019}.
Conveniently, $\tau_p$ also directly probes correlations in time (which the steady-state averages of $S_{L/2}$ and $I_{AB}$ do not).
The qualitative physics of $\tau_p$ can be simply understood by considering the fate of an initially localized bit of entropy (e.g. a single maximally mixed qubit).
For high measurement rates, this bit of entropy remains localized and hence $\tau_p$ is independent of system size and approaches a constant in the thermodynamic limit. Meanwhile, at low measurement rates, this bit of entropy becomes delocalized and is unlikely to purify, so $\tau_p$ diverges with system size.
At the critical point, we expect $\tau_p \sim L^z$, where $z$ is the dynamical exponent, making it a particularly convenient observable for finite-size scaling.
Finally, we complement our study of the median purification time by investigating the global entropy of an initially maximally mixed state as a function of time, $S(t)$ \cite{gullans_dynamical_2020}; indeed, $\tau_p$ can be understood simply as the half life of $S(t)$.
For high measurement rates, $S(t)$ decays exponentially, while for low measurement rates, $S(t)$ becomes time-independent.

\emph{Long-range interactions  with $\alpha\gtrsim 3$.---}As a starting point for our analysis, let us consider fixed $\alpha=3.5$.
The observables we investigate exhibit clear evidence of an entanglement phase transition at a critical measurement rate, $p_c$ [Fig.~\ref{fig:3}(a-c)].
Perhaps the most striking signature of the transition comes from the anti-podal mutual information, which exhibits a peak at the critical point that sharpens with increasing system size [Fig.~\ref{fig:3}(a)]. 
Moreover, the height and location of this peak are independent of $L$, consistent with prior observations in short-range-interacting hybrid  circuits~\cite{li_measurement-driven_2019}. 
This is a consequence of conformal symmetry at the critical point~\cite{li_measurement-driven_2019}, and suggests that the measurement-induced transition remains a CFT  for sufficiently weak power-laws (Fig.~\ref{fig:2}). 
To quantitatively characterize the transition,  we perform finite-size scaling [inset, Fig.~\ref{fig:3}(a)] of $I_\mathrm{AB}$ using the scaling form:
\begin{equation}
    I_\mathrm{AB} = L^\beta f((p - p_c)L^{1/\nu}).
\end{equation}
Crucially, this allows us to extract both the scaling dimension, $\beta$, of $I_\mathrm{AB}$ and the correlation length exponent $\nu$.
We find, $\beta \approx 0$ and $\nu \approx 1.3$ (Fig.~\ref{fig:2}), consistent with all prior results  in short-range interacting models~\cite{skinner_measurement-induced_2019, li_measurement-driven_2019, gullans_dynamical_2020, zabalo_critical_2020, 3_foot_cft_beta} 
%\footnote{In fact, in the short-range case $\beta=0$ is fixed by conformal symmetry in space-time, but we do not assume any symmetry here.}. % add error bars. 

In order to extract the dynamical critical exponent, we turn to an exploration of  the median purification time, $\tau_p$. 
As shown in Fig.~\ref{fig:3}(b), we observe a \emph{single} crossing point (which independently identifies $p_c$) for $\tau_p/L$ across all system sizes. This is  consistent with the dynamical scaling hypothesis,
\begin{equation}
    \tau_p(p) = L^z g[\,(p - p_c) L^{1/\nu}\,],
\end{equation}
with $z=1$ (as expected for a CFT).
The conformal nature of the transition is further confirmed by the finite-size-scaling collapse depicted in the inset of Fig.~\ref{fig:3}(b).
A few remarks are in order.
First, we find that the correlation length exponent extracted from $\tau_p$ gives $\nu \approx 1.3$, in excellent agreement with both with the short-range transition and the scaling analysis of $I_\mathrm{AB}$~\cite{supp}.
Second, one hopes that the critical exponents extracted from $\tau_p$ can be used to directly collapse the full time dynamics of the global entropy, $S(t)$. 
This is indeed born out by the data [Fig.~\ref{fig:3}(c)], where we have utilized the general scaling form,
\begin{equation}
    S(p, t) = h(\,(p - p_c) L^{1/\nu},\, t/L^z\,).
\end{equation}

Although we have focused our discussions on the specific case of $\alpha=3.5$, an extensive numerical study (see supplementary materials) of the transition for all $\alpha\gtrsim 3$ reveals the same physics~\cite{supp}.
In particular, the critical exponents $\nu$ and $z$ are found to agree with their short-range values, implying that the universality class of the measurement-induced transition is unchanged for $\alpha\gtrsim 3$.

\emph{Long-range interactions  with $\alpha \lesssim 3$.---}We now turn our attention toward understanding the new physics that arises for $\alpha\lesssim3$.
To be concrete, let us begin by applying the same diagnostic toolset to long-range hybrid circuits with $\alpha=2.25$.  
Two profound differences  emerge: (i) the location and height of the peak of $I_\mathrm{AB}$ drifts with system size [Fig.~\ref{fig:3}(d)], and (ii) $\tau_p$ no longer exhibits a single crossing point [Fig.~\ref{fig:3}(e)].
These trends immediately imply $\beta \neq 0$ and $z \neq 1$, indicating that sufficiently strong power-laws alter the universality class of the transition.
More specifically, the critical point is no longer described by a CFT. 

To determine precisely when the universality class of the transition changes, we extract $\nu(\alpha)$ and $z(\alpha)$ via the purification time and the collapse of $S(t)$ [Fig.~\ref{fig:3}(e,f)]~\cite{4_foot_taup_v_iab}.
%\footnote{We use $\tau_p$ rather than $I_\mathrm{AB}$ because the latter is an inaccurate quantitative probe of the transition for $\alpha<3$; see~\cite{supp}.}. 
%
As shown in Fig.~\ref{fig:2}, for $\alpha \lesssim 3$, we find that $\nu$ and $z$ vary continuously; this identifies $\alpha \approx 3$ as the threshold for which long-range effects become relevant for the measurement-induced transition. % error bars. Find alpha interval where z=1 within errors.

Interestingly, further reducing $\alpha$ yields additional modifications to the transition.
Specifically, despite clear evidence for the presence of an entanglement transition, we find that for $\alpha<2$ the half-chain entanglement entropy always scales with system size even at very high measurement rate, i.e., there is no longer a true area-law phase.
Instead, there is a ``sub-volume" law phase, where $S_{L/2} \sim L^\mu$, with $0 < \mu < 1$~\cite{supp}.
The emergence of this sub-volume law phase can be understood quite simply by analyzing the short-time half-chain entanglement generated by our dynamics.
Indeed, a single layer of long-range gates contributes additional entropy $\sim L^{2-\alpha}$ for $\alpha<2$.
Numerical analysis indicates this bound is approximately tight, and we conjecture $\mu = 2-\alpha$~\cite{supp}.

At $\alpha=1$, the ``sub-volume" scaling becomes a true volume law, and the half-chain entanglement is no longer a valid indicator of the measurement-induced transition.
However, observables that are not geometrically local, such as $S(t)$, $\tau_p$ and the entangling power~\cite{vijay_measurement-driven_2020}, do not suffer from this limitation and demonstrate that a measurement-induced transition still occurs at $\alpha=1$ (and indeed, even in all-to-all connected models~\cite{vijay_measurement-driven_2020, nahum_measurement_2020}). 

\emph{Effective quantum spin model.---}To provide a theoretical understanding for the change of universality at $\alpha \approx 3$, we develop a mapping that relates the steady-state entanglement entropy of our long-range hybrid quantum circuit to the ground-state properties of a long-range 1D quantum Ising model~\cite{supp,bao2021symmetry}. 
This mapping hinges on a conditional R\'enyi entropy (which is related to $S_{L/2}$ via the replica method~\cite{hayden2016holographic,nahum_operator_2018}), 
\begin{equation}
	S_A^{(2)} = -\log\left(\overline{ \sum_{m} p_{m}^2 \tr\rho_{A,m}^2 }\right) + \log\left(\overline{ \sum_{m} p_{m}^2 } \right),
\end{equation}
where $\overline{\hspace{1mm}  \boldsymbol{\cdot}   \hspace{1mm}}$ represents an average over circuit realizations, and $\rho_{A,m}$ is the reduced density matrix for subsystem $A$, conditioned on a specific set of measurement outcomes, $m$, with probabilities $p_{m}$~\cite{supp,bao2021symmetry}.
Much like the half-chain entanglement entropy, $S_A^{(2)}$ undergoes an area- to volume-law transition as a function of the measurement rate~\cite{bao_theory_2020, jian2020measurement}.
Crucially, although this transition belongs to a different universality class, it is analytically tractable and will provide  insights into the original transition.

In order to compute $S_A^{(2)}$, we consider a slightly modified circuit that trades random \emph{connectivity} for random \emph{interaction strengths}. 
%
% We emphasize this modification  the essential scrambling and measurement dynamics of this circuit are identical to that considered in previous sections (Fig.~\ref{fig:1}). 
%
To be precise, we consider a circuit consisting of layers of single-qubit Haar random unitaries, projective local-Z measurements, and long-range Ising interactions $\theta_{ij} Z_i Z_j$, where $\theta_{ij}$ are drawn from a Gaussian distribution with zero mean and variance $\propto 1/\abs{i-j}^\alpha$.
The scrambling properties of this circuit are similar to those in our original long-range  circuit [Fig.~\ref{fig:1}(a)], and we believe that it undergoes a measurement-induced transition of the same universality class (as long as one considers the same observable). 

Somewhat remarkably, one can compute $S_A^{(2)}$ for the modified circuit, via an exact mapping to imaginary time evolution under a long-ranged Ising Hamiltonian \cite{supp,bao2021symmetry}: 
\begin{equation}
    H_{\rm eff} = - \sum_{ij} \frac{J}{\abs{i-j}^\alpha} \left( 3 \sigma^z_i \sigma^z_j  - \sigma^x_i \sigma^x_j \right) -\sum_i h \sigma^{x}_i.
\end{equation}
In this context, the measurement-induced transition in $S_A^{(2)}$ can be understood as the symmetry-breaking Ising transition in the ground state of $H_{\rm eff}$~\cite{bao2021symmetry}.

To this end, let us recall the effect of long-range interactions on the universality class of the  Ising transition. 
In particular, one can consider the long-range tail as a perturbation to the  action of the short-ranged model, $\delta S= \int dq \, d \omega \, q^{\alpha-1}\phi_q \phi_{-q}$, where $q$ is the momentum, $\omega$ is the Matsubara frequency  and $\phi$ is the  order parameter~\cite{sak_recursion_1973, fisher1972critical}.
At the (short-ranged) Ising critical point, the scaling dimension of  $\delta S$ is  $3-\alpha-\eta$, where $\eta/2$ is the scaling dimension of the order parameter.
Thus, the long-range coupling becomes a relevant perturbation for the Ising transition when $\alpha<3-\eta$.  
This insight immediately allows us to understand why the measurement-induced transition's universality changes at  $\alpha \lesssim 3$.

More precisely, we also expect long-range interactions to become relevant for the measurement-induced transition at $\alpha = 3-\eta$, where $\eta$ is now the anomalous dimension of the short-range measurement-induced transition.
%in the modified circuit model from this section, η = 1/4, (ii) in a hybrid circuit with infinite qudit dimension, the transition is described by a perco- lation critical point and η = 5/24, (iii) in numerics on short-range interacting hybrid Clifford circuits, η ≈ 0.22 [26]. 
Although difficult to compute directly, one can estimate $\eta$ in three ways: (i) in the modified circuit model from this section, the transition of $S_A^{(2)}$ has $\eta=1/4$, (ii) in a Haar-random hybrid circuit with infinite qudit dimension, the transition of $S_{L/2}$ has $\eta=5/24$ (and is described by percolation), (iii) in numerics on short-range interacting hybrid Clifford circuits, one finds $\eta \approx 0.22$ \cite{zabalo_critical_2020}. 
%This adjustment, known as the Sak's correction \cite{sak_recursion_1973}, is generic in long-range quantum roter models \cite{defenu_criticality_2017}, and can be somewhat large (e.g. $1/4$ in the $d=1$ Ising model).
%
All of these calculations suggest  $\alpha \approx 2.8$ as the critical threshold for the relevance of long-range interactions, consistent with our numerical phase diagram [Fig.~\ref{fig:2}].
%
%Carrying out a more thorough analysis to accurately identify the correction to $\alpha_0=3$ remains a direction for future research. 

Our work opens the door to a number of intriguing future directions.
First, it is natural to conjecture a connection between scrambling light-cones in power-law interacting systems and the universality class of the measurement-induced transition in long-range hybrid circuits.
Sharpening this connection could  provide a quantum information theoretic understanding of our phase diagram. 
Second, our predicted phase diagram can be directly probed in current generation trapped ion experiments, where the long-range interaction can, in principle, be tuned between $0 < \alpha < 3$ \cite{porras_effective_2004,pagano2020quantum}; however, developing a scalable experimental probe of the transition remains an important challenge~\cite{gullans_scalable_2019}.

\emph{Note added:} During the completion of this work, we
became aware of complementary work on the measurement induced transition in long-range interacting Hamiltonian systems~\cite{minato_fate_2021}.

%During the preparation of this manuscript we became aware of a related work \cite{minato_fate_2021}.

\emph{Acknowledgements.---}We gratefully acknowledge the insights of and discussions with Michael Gullans, Joel Moore, and Adam Nahum. We are particularly indebted to Chao-Ming Jian for pointing out the anomalous dimension for hybrid circuits with infinite qudit dimension. 
This work was supported by the U.S. Department of Energy, Office of Science, Office of Advanced Scientific Computing Research, under the Accelerated Research in Quantum Computing (ARQC) program and by the U.S. Department of Energy, Office of Science, National Quantum Information Science Research Centers, Quantum Systems Accelerator (QSA).
MB acknowledges support  through the Department of Defense (DoD) through the National Defense Science \& Engineering Graduate (NDSEG) Fellowship Program. SC acknowledges support from the Miller Institute for Basic Research in Science. EA is supported in part by the Department of Energy project DE-SC0019380 ``The Geometry and Flow of Quantum Information: From Quantum Gravity to Quantum Technology" and by the Gyorgy Chair in Physics at UC Berkeley.

\bibliographystyle{apsrev4-2}
\bibliography{bibliography}

\begin{thebibliography}{71}
\bibitem [{3_f()}]{3_foot_cft_beta}%
  \BibitemOpen
  \href@noop {} {}\bibinfo {note} {In fact, in the short-range case $\beta=0$
  is fixed by conformal symmetry in space-time, but we do not assume any
  symmetry here.}\BibitemShut {Stop}%
\bibitem [{4_f()}]{4_foot_taup_v_iab}%
  \BibitemOpen
  \href@noop {} {}\bibinfo {note} {We use $\tau_p$ rather than $I_\mathrm{AB}$
  because the latter is an inaccurate quantitative probe of the transition for
  $\alpha<3$; see~\cite{supp}.}\BibitemShut {Stop}%
\end{thebibliography}

\end{document}

\title{Supplementary Material: The Measurement-induced Transition in  Long-range Interacting Quantum Circuits}

\author{Maxwell Block}
\affiliation{Department of Physics, University of California, Berkeley, California 94720, USA}

\author{Yimu Bao}
\affiliation{Department of Physics, University of California, Berkeley, California 94720, USA}

\author{Soonwon Choi}
\affiliation{Department of Physics, University of California, Berkeley, California 94720, USA}
\affiliation{Center for Theoretical Physics, Massachusetts Institute of Technology, Cambridge, MA 02139, USA}

\author{Ehud Altman}
\affiliation{Department of Physics, University of California, Berkeley, California 94720, USA}
\affiliation{Materials Sciences Division, Lawrence Berkeley National Laboratory, Berkeley, California 94720, USA}

\author{Norman Y. Yao}
\affiliation{Department of Physics, University of California, Berkeley, California 94720, USA}
\affiliation{Materials Sciences Division, Lawrence Berkeley National Laboratory, Berkeley, California 94720, USA}

\date{\today}

\maketitle
\tableofcontents

\section{Numerical Methods}
In this section, we provide details on the methods used to simulate our long-range hybrid circuit model and perform finite-size scaling analysis.
As discussed in the main text, we use four observables to examine the transition: half-chain entanglement entropy ($S_{L/2}$), anti-podal mutual information ($I_{AB}$), single qubit purification time ($\tau_p$), and global entropy ($S$).
We obtain numerical estimates of these observables through several simulations, summarized in Table~\ref{tab:observable-sims}.
For the antipodal mutual information, purification time, and global entropy, we additionally perform finite size scaling to extract critical exponents, the results of which are shown in Sec.~\ref{sec:fss-summary}.
As mentioned in the main text, all simulations are based on \emph{Clifford} dynamics.
Below, we give details regarding the analysis of each observable that augment the discussion in the main text.

\renewcommand{\arraystretch}{1.5}
\begin{table*}
\centering
\begin{tabular}{c|c|c|c|c|c}
\hline \hline
 \makecell{Observables \\ } & \makecell{System Sizes \\ $L$} & \makecell{Power Laws  \\ $\alpha$} & \makecell{$p$ Resolution \\ $\Delta p$} & \makecell{Circuit Depth \\ $D$} & \makecell{Circuit Samples \\ $n$} \\
 \hline
 $S$, $I_{AB}$, $S_{L/2}$ & [16, 32, ... 512] & [2.0, 2.25, ..., 4.0] & 0.004 & $32 L$ & 300 \\
 \hline
 $S$, $S_{L/2}$ & [16, 32, ... 512] & [1.0, 1.25, ..., 2.0] & 0.01 & $32 L$ & 300 \\
 \hline
 $\tau_p$ & [16, 32, ... 512] & [1.0, 1.25, ... 4.0] & 0.005 & $16 L$ & 700 \\
 \hline
 $\tau_p$ & [16, 32, ... 1024] & [2.8, 2.83, ... 3.1] & 0.002 & $16 L$ & 800 \\
 \hline
 $S_{L/2}$ & [16, 32, ... 1024] & [1.5, 3.5] & $p = [0.05, 0.5, 0.75]$ & $4L$ & 500\\
 \hline
 \hline
\end{tabular}
\caption{Summary of simulations. Circuit depth is defined as the number of gates in the circuit \emph{divided} by $L/2$, so there are $L/2$ gates/time step. Note that $\Delta p$, $D$, and $n$ are all chosen according to the accuracy required for that observable.In particular, purification time, is crucial for our quantitative analysis and hence we choose $\Delta p$ to be small to in those simulations.}
\label{tab:observable-sims}
\end{table*}

\subsection{Half-chain Entanglement Entropy} \label{sec:hcee}

The half-chain entanglement entropy was one of the first observables used to diagnose measurement induced criticality in short-range models \cite{li_quantum_2018, skinner_measurement-induced_2019}.
However, it is of limited in use for probing the universality of the transition in long-range models for the simple reason that it requires a notion of geometric locality to be well defined, which breaks down as $\alpha \rightarrow 0$. 

\begin{figure}[h!]
	\centering
	\includegraphics[width=1.0\linewidth]{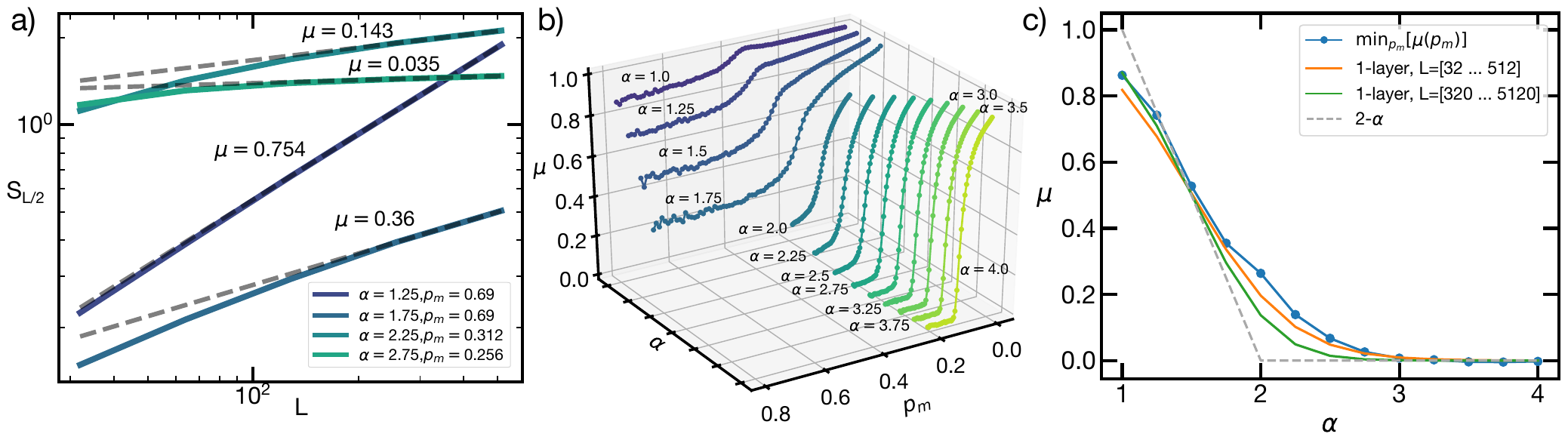}
	\caption{(a) Selected $S_{L/2}(L)$ with power law fits. The fit for $\alpha=1.25, p_m=0.69$ clearly shows a consistent scaling $S_{L/2} \sim L^0.754$, i.e. a sub-volume law phase. (b) The $S_{L/2}(L)$ scaling power $\mu$ as a function of $\alpha$ and $p$. For $\alpha < 2$, sub-volume law scaling is observed for high measurement rates. (c) Numerical single-layer model of $\mu$ compared to $\mu$ estimate from power-law fits. The model applied to the same system sizes as the data (orange) demonstrates reasonable agreement with the estimated $\mu$. The model also shows the transition in $\mu$ sharpens as system size is increased (green), approaching $2-\alpha$ in the thermodynamic limit.}
	\label{fig:subvol}
\end{figure}

Nonetheless, the degenerating behavior of $S_{L/2}$ as $\alpha \rightarrow 0$ is itself a indication of the changing universality of the transition which is worth understanding in its own right.
To get some intuition for how $S_{L/2}$ behaves as $\alpha$ is decreased, we consider the high-measurement rate limit where the entire system is frequently returned to a near-product state.
In this case, the entanglement entropy generated by the dynamics cannot accumulate, and will be dominated by whatever entanglement is generated at short times, i.e. in a single layer of the circuit.
This short time entanglement, $\Delta S_{\rm instant}$, will be proportional to the number of gates crossing between the two chain halves, and hence we expect
\begin{align}
    \Delta S_{\rm instant} &\propto \int_1^{L/2} \; dx \; \int_{L/2+1}^{L} \; dy \; \frac{1}{\abs{x-y}^{\alpha}} \\
    &\propto L^{2-\alpha} + O(1) \text{ as } L \rightarrow \infty.
\end{align}
We examine this hypothesis quantitatively by simulating $S_{L/2}$ for various $L$ (Tab.~\ref{tab:observable-sims}) and fitting the result to a power law, $S_{L/2} \approx A L^\mu$.
As shown in Fig.~\ref{fig:subvol}(a), this power-law fit is typically quite good for larger system sizes.
We can then consider how $\mu$ depends on $\alpha, p$, which is shown in Fig.~\ref{fig:subvol}(b).
As expected, for $\alpha \gg 2$ we have $\mu \approx 0$ above some critical measurement right, with $\mu$ rising rapidly to $1$ below this rate [Fig.~\ref{fig:subvol}(b)].
For $\alpha \ll 2$ the behavior is starkly different; $\mu > 0$ even for large measurement rates, and $\mu \approx 1$ again for small measurement rates [Fig.~\ref{fig:subvol}(b)].

To evaluate how well the short-time approximation described above explains these effects, we attempt to model the \emph{minimum} $\mu$ determined for each $\alpha$.
To account for finite size effects, we use a simple numerical routine to count the number of single-layer crossings rather than the analytic formula above.
The result is in fairly good agreement with the extracted $\mu$, and satisfactorily demonstrates the smooth change in $\mu$ is a consequence of finite-size effects [Fig.~\ref{fig:subvol}(c)].
We observe this single-layer model yields a sharper transition for $\mu$ if it is run on larger-system sizes [Fig.~\ref{fig:subvol}(c)], indicating that $\mu(\alpha) \rightarrow 2-\alpha$ in the thermodynamic limit.

\subsection{Anti-podal Mutual Information}

The anti-podal mutual information is a convenient observable for careful extraction of critical exponents in short-range systems owing to its sharp peak around the critical measurement rate, which strongly constrains the correlation length exponent $\nu$ (Fig.~\ref{fig:ap-mi-raw}) \cite{li_measurement-driven_2019}. 
In short range-systems, conformal symmetry enforces the scaling form
\begin{equation} \label{eq:iab-cft}
    I_{AB} = f((p-p_c)L^{1/\nu}).
\end{equation}
In long-range systems however, where conformal symmetry may be absent, the scaling dimension of $I_{AB}$ need not be $0$ and we augment this scaling form to be
\begin{equation} \label{eq:iab-simple}
    I_{AB} = L^\beta f((p-p_c)L^{1/\nu}).
\end{equation}
The necessity of introducing the prefactor can be observed the $I_{AB}$ data for $\alpha < 3$ (Fig.~\ref{fig:ap-mi-raw}).

\begin{figure}[h!]
	\centering
	\includegraphics[width=1.0\linewidth]{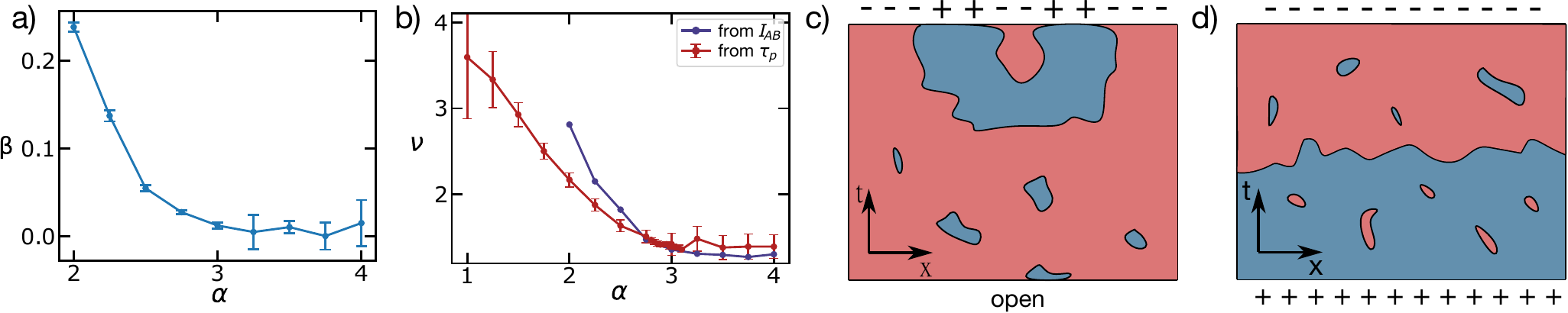}
	\caption{(a) The scaling dimension of $I_{AB}$, $\beta$, as a function of power-law $\alpha$. For $\alpha>3$, we find $\beta \approx 0$ as expected from CFT; for $\alpha<3$, $\beta$ increases dramatically, demonstrating the altered universality of the transition. (b) A comparison of $\nu$ as extracted from $I_{AB}$ and $\tau_p$. The estimates disagree, which we attribute to the highly anisotropic domain walls probed by $I_{AB}$. (c) A typical domain in the ordered phase of the spin model associated with our dynamics with $I_{AB}$ boundary conditions. (d) The analogous domain for $S$ boundary conditions.}
	\label{fig:nu-compare}
\end{figure}

Like $S_{L/2}$, $I_{AB}$ relies on a notion of geometric locality that breaks down as $\alpha \rightarrow 0$. 
Indeed, the same short-time analysis described in Sec.~\ref{sec:hcee} shows that for $\alpha<2$ $I_{AB}$ can scale strongly with system size even under large measurement rates -- the onset of this behavior can be seen for $\alpha = 2.0$ in Fig.~\ref{fig:ap-mi-raw}.
However, $I_{AB}$ is a more nuanced probe of the transition than $S_{L/2}$, and indicates the change in universality at $\alpha=3$ in two ways.
First, it is precisely for $\alpha \leq 3$ that we find $\beta > 0$, which is directly tied to the lack of conformal symmetry in this regime [Fig.~\ref{fig:nu-compare}(a)].
Second, and more intriguingly, the correlation length exponent $\nu$ we extract from collapsing $I_{AB}$ actually disagrees with the $\nu$ we extract from collapsing $\tau_p$ for $\alpha<3$ [Fig.~\ref{fig:nu-compare}(b)].

To see the origin of this discrepancy, consider the horseshoe-shaped domain favored by $I_{AB}$ boundary conditions in the ordered phase of the associated spin model~\cite{bao_theory_2020,jian2020measurement} [Fig.~\ref{fig:nu-compare}(b)].
The domain is characterized by a separation along the top boundary and a penetration depth $T_{AB}$ back into the bulk, as shown in Fig.~\ref{fig:nu-compare}(c).
The $I_{AB}$ scaling function can thus depend on two dimensionless quantities
\begin{equation} \label{eq:iab-full}
    I_{AB} = L^\beta \tilde{f}(L/\xi_x, T_{AB}/\xi_t),
\end{equation}
where $\xi_x, \xi_t$ are the space and time correlation lengths respectively.
In short-range models, conformal symmetry demands that $\xi_t=\xi_x$ and $T_{AB} \propto L$, so these arguments are redundant leaving us with the scaling form of Eq.~\eqref{eq:iab-cft}.
In the long-range regime ($\alpha \lesssim 3$), the conformal symmetry is absent, and we expect $T_{AB} \propto L^y$. 
The scaling function in Eq.~\eqref{eq:iab-full} reduces to Eq.~\eqref{eq:iab-simple} \emph{only} if $y=z$.
Therefore, there is no reason to think that Eq.~\eqref{eq:iab-simple} will generally collapse $I_{AB}$, much less enable one to extract $\nu$.
Indeed, the fact Eq.~\eqref{eq:iab-simple} \emph{does} collapse the data so well is somewhat mysterious (Fig.~\ref{fig:ap-mi-collapse}), although it is expected that the $\nu$ extracted from this collapse is incorrect and deviates from that extracted from $\tau_p$ [Fig.~\ref{fig:nu-compare}(b)].

\subsection{Single Qubit Purification Time}

As described in the main text, to extract they dynamical exponent $z$ we require a new observable, which we call the single-qubit purification time $\tau_p$.
It is closely related to a probe of the transition suggested in \cite{gullans_scalable_2019}.
Quite simply, $\tau_p$ is the median time it takes for a single maximally mixed qubit purify under our hybrid dynamics.
However, a product state with a single maximally qubit could be purified at early times due to a ``lucky" measurement with probability $O(1/L)$.
To eliminate this finite-size effect, we apply a globally scrambling Clifford operation to the initial state [Fig.~3(b)].
Then, we expect $\tau_p$ to probe the bulk correlation time, and have the scaling form,
\begin{equation}
    \tau_p(p) = L^z f((p - p_c) \cdot L^{1/\nu}).
\end{equation}

To assign errors to $z, \nu$ extracted from collapsing $\tau_p$ (Fig.~\ref{fig:ptime-collapse}), we employ a standard bootstrap analysis.
Of the 700 (800) samples of purification time, we sub-sample 600 (700) with replacement, and define compute $\bar{\tau_p}$ as the median of these samples.
We then optimize the collapse of $\bar{\tau_p}$, to obtain $\bar{z}, \bar{\nu}$.
Repeating this procedure for $2500$ bootstrap sub-samples, we obtain distributions of $\bar{z}, \bar{\nu}$ that enable us to determine confidence intervals for these quantities.
The error-bars shown in Fig.~2 denote $95\%$ confidence intervals.

\subsection{Global Entropy}

As described in the main text, we primarily use the global entropy, $S$, to corroborate our analysis of $\tau_p$.
Because $S$ cannot scale with system size in the purifying phase, the scaling form is constrained to be
\begin{equation}
    S(p, t) = h((p - p_c) \cdot L^{1/\nu}, t/L^z).
\end{equation}
That is, $S$ can depend on both correlation lengths but does \emph{not} have unknown scaling dimension.
This makes it an excellent test for the critical exponents extracted from $\tau_p$, since no additional fitting is required and the quality of collapse reflects on the accuracy of those exponents.
We find $S$ is collapsed extremely well by the $\tau_p$ exponents across all $\alpha$ we studied (Fig.~\ref{fig:global-s-collapse}).

\section{Effective quantum Hamiltonian for long-range quantum circuits}
In this section, we map the entanglement entropy in the steady state of our hybrid quantum circuit to the properties of the \emph{ground state} of a one-dimensional quantum Ising model.
This allows us to understand the change in universality of the transition from the perspective of conventional effective field theory.
In particular, the scaling analysis of the effective quantum Ising model predicts a change in universality at $\alpha \approx 3$, consistent with the numerical simulation presented in the main text.
We elaborate on the details of this mapping below.

% \subsection{Long-range quantum circuit}
\begin{figure}[t!]
	\centering
	\includegraphics[width=0.7\linewidth]{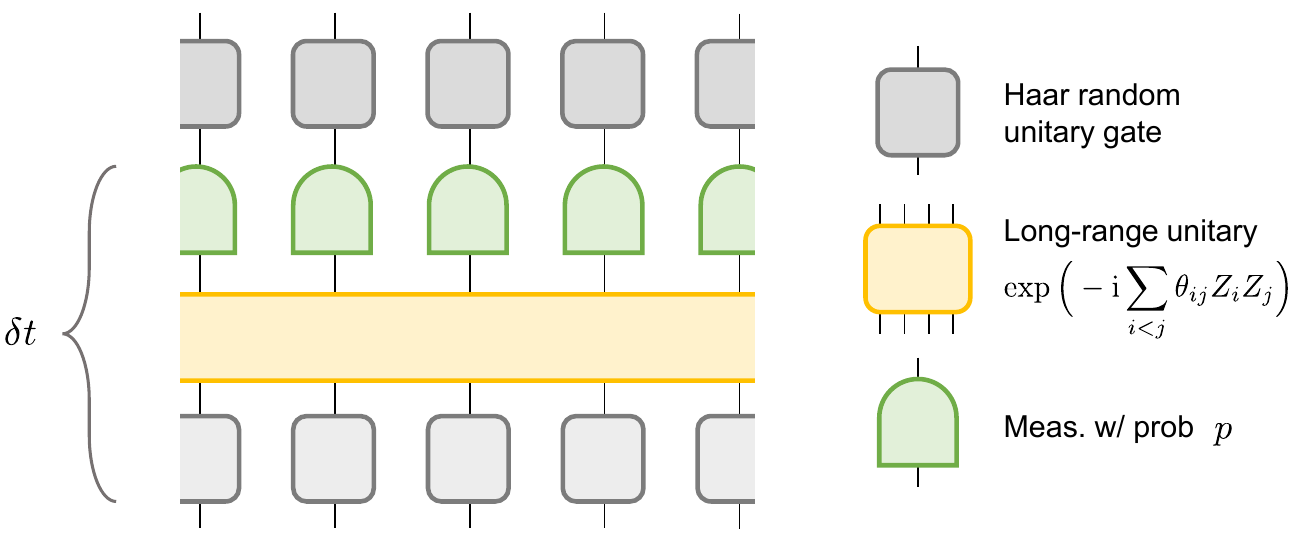}
	\caption{Schematic of the hybrid-circuit model whose dynamics we map to imaginary time evolution under an effective Hamiltonian. The dynamics consists of Haar random unitaries followed by random long-range Ising interactions and projective measurements.}
	\label{fig:circuit}
\end{figure}

We consider a quantum circuit operating on a one-dimensional array of $L$ qubits. 
The circuit consists of unitary gates and measurements arranged in a layered structure as shown in Fig.~\ref{fig:circuit}.
Within each time step, we first apply single-qubit Haar-random unitary gates at every site.
We then apply a long-range Ising gate
\begin{align}
    U_{\text{LR}} = \exp\left(-\mathrm{i}\sum_{1 \leq i < j \leq L} \theta_{ij} Z_i Z_j\right),
\end{align}
where $Z_i$ is the Pauli-Z operator acting on the $i^\mathrm{th}$ qubit, and $\theta_{ij}$ are \emph{independent} Gaussian random variables with zero mean and variance $K/|i-j|^\alpha$. 
%
%We note that Gaussian random coupling is chosen for simplicity, one can map the dynamics to an effective ground state problem as long as the distribution of the random coupling is symmetric to zero.
%
Finally, to conclude a time step, we perform projective measurements in the computational basis on every qubit with probability $p$.
The circuit consists of $N_T$ time steps. 
%
% We note that the random couplings in different gates are drawn independently from their own distributions.
%
% We are interested in the limit of $L, N_T \gg 1 $.

% \subsection{Effective quantum Hamiltonian}
In principle, we wish to study the entanglement phase transition that can be detected by the von Neumann entropy of the subsystem $A$,
\begin{align}
    S_A = \overline{\sum_m p_m S_{A,m}} = \overline{\sum_m -p_m\tr\left(\rho_{A,m}\log\rho_{A,m}\right)},
\end{align}
where the overline $\overline{\cdot}$ represents the average over random unitary gates and $p_m$ is the probability of particular trajectory (set of mesaurement outcomes).
We use the framework developed in Ref.~\cite{bao2021symmetry} to  express the von Neumann entanglement entropy as a limit of the ``conditional R\'enyi entropy" of order $n$,
\begin{align}
    S^{(n)}_A &= \frac{1}{1-n}\log\left(\,\overline{\tr\rho_{MA}^n}\,\right)-\frac{1}{1-n}\log\left(\,\overline{\tr\rho_{M}^n}\,\right)=\frac{1}{1-n} \log \left(\frac{\overline{\sum_m p_m^n \tr ( \rho_{A,m}^n )} }{\overline{\sum_m p_m^n}}
\right).
\end{align}
which recovers the von Neumann entropy in the replica limit $n \to 1$.
As shown in Ref.~\cite{bao2021symmetry}, the $n^{\text{th}}$ conditional R\'enyi entropy can be studied in the duplicated Hilbert space $\mathcal{H}^{(n)} = (\mathcal{H}\otimes \mathcal{H}^*)^{\otimes n}$ and expressed in term of a matrix element
\begin{align}
    S^{(n)}_A = \frac{1}{1-n}\log\left(\frac{\bbrakket{\mathcal{I}}{\mathcal{C}_{\ell,A}^{(n)}|\tilde{\rho}^{(n)}}}{\bbrakket{\mathcal{I}}{\tilde{\rho}^{(n)}}}\right),
    \label{eq:nth_entropy_eff}
\end{align}
where $\kket{\tilde{\rho}^{(n)}} \equiv \sum_m \kket{\rho_m^{(n)}}$ is a vector in $\mathcal{H}^{(n)}$, and $\kket{\rho_m^{(n)}} \equiv \tilde{\rho}_m^{\otimes n}$ represents $n$ copies of the un-normalized density matrix in the trajectory with $m$.
Similarly, $\kket{\mathcal{I}}$ is the identity operator (vectorized in $\mathcal{H}^{(n)}$), so $\bbrakket{\mathcal{I}}{\rho_m^{(n)}} = \tr (\tilde{\rho}_m^n) = p_m^n$.
Finally, the operator $\mathcal{C}_{\ell,A}^{(n)}$ cyclically permutes between the replicated Hilbert spaces on sites in $A$.
Formally,
\begin{align}
    \mathcal{C}_{\ell,A}^{(k)} &= \sum_{\{a_i\}}  \bigotimes_{i=1}^k \left( \ket{a_{i+1}}\bra{a_{i}} \otimes \mathds{1} \right),
\end{align}
where $a_i$ runs over the basis state of the subsystem $A$.
This cyclic permutation operator allows the calculation of the $n^{\text{th}}$ moment $p_m^n\tr\rho_{A,m}^n = \bbrakket{\mathcal{I}}{\mathcal{C}_{\ell,A}^{(n)}|\tilde{\rho}^{(n)}}$.
The replica method for the subsystem entanglement entropy can be generealized to study other probes considered in the paper, such as the purification entropy and the mutual information.
The generalization is straightforward by considering the matrix element with $\mathcal{C}_{\ell,A}^{(n)}$ operating on different regions and different initial states of the circuits \cite{bao2021symmetry}.

Having established our mathematical framework, we now focus on the specifics of our long-range circuit dynamics (Fig.~\ref{fig:circuit}).
The time evolution of the replicated density matrix $\kket{\tilde{\rho}^{(n)}}$ in $\mathcal{H}^{(n)}$ is given by
\begin{align} \label{eq:evolution}
    \kket{\rho^{(n)}(t+\delta t)} = \mathcal{M}^{(n)}\mathcal{U}_{\text{LR}}^{(n)}\mathcal{U}_{\text{Haar}}^{(n)}\kket{\rho^{(n)}(t)}.
\end{align}
where $\mathcal{U}_{\text{Haar}}^{(n)}$, $\mathcal{U}_{\text{LR}}^{(n)}$, and $\mathcal{M}^{(n)}$ denote the averaged Haar random gates, long-range gates, and measurements in $\mathcal{H}^{(n)}$, respectively.
In Ref.~\cite{bao2021symmetry}, it is shown that the evolution of $\kket{\rho^{(n)}}$ is quite generally described by an imaginary time evolution under an effective quantum Hamiltonian.
In the following, we will focus on the case of $n = 2$ and explicitly derive the effective Hamiltonian for our circuit.
For convenience, we drop the superindex $(n)$.

First, we consider the averaged Haar random unitary gates in the duplicated Hilbert space $\mathcal{H}^{(2)}$, $\mathcal{U}_{\text{Haar}} = \bigotimes_{j = 1}^L \mathcal{U}_{\text{Haar},j}$ with $\mathcal{U}_{\text{Haar},j} = \overline{U \otimes U^* \otimes U \otimes U^*}$.
For Haar random unitary gates, the averaging can be exactly carried out
\begin{align}
    \mathcal{U}_{\text{Haar},j} = \sum_{\sigma_j,\tau_j \in \{\mathcal{I}, \mathcal{C}\}} w_g(\tau_j,\eta_j)\kket{\tau_j}\bbra{\eta_j}_j,
\end{align}
where the Weingarten function $w_g(\tau_j,\eta_j) = \delta_{\tau_j\eta_j}/2 - 1/6$~\cite{collins2003moments}, $\tau_j$ and $\eta_j$ are either the identity $\mathcal{I}$ or the swap operation $\mathcal{C}$ in the permutation group $\mathcal{S}_2$, and $\kket{\tau_j}_j$ is given by
\begin{align}
    \kket{\tau_j}_j = \left\{\begin{array}{cc}
    \sum_{a,b = \pm 1} \kket{aabb}_j & \tau_j = \mathcal{I} \\
    \sum_{a,b = \pm 1} \kket{abba}_j & \tau_j = \mathcal{C}
    \end{array}\right.,
\end{align}
where $\kket{aabb}_j$ and $\kket{abba}_j$ are computational basis states of the duplicated Hilbert space $\mathcal{H}^{(2)}$ at site $j$.
The averaged unitary gate facilitates the projection onto an effective two-dimensional local Hilbert space spanned by $\kket{\mathcal{I}}_j$ and $\kket{\mathcal{C}}_j$.
Here, it's convenient to work in an orthonormal basis of the effective Hilbert space:
\begin{align}
    % \kket{\uparrow}_j &= \frac{\sqrt{3}+1}{2\sqrt{6}}\kket{\mathcal{I}}_j - \frac{\sqrt{3}-1}{2\sqrt{6}}\kket{\mathcal{C}}_j, \\
    \kket{+}_j &= \left(\kket{\mathcal{I}}_j + \kket{\mathcal{C}}_j\right)/(2\sqrt{3}) \\
    \kket{-}_j &= \left(\kket{\mathcal{I}}_j - \kket{\mathcal{C}}_j\right)/2.
\end{align}
We also define the usual Pauli operators to act on this space such that $\sigma_j^x$ has eigenstates $\kket{\pm}_j = (\kket{\uparrow}_j\pm\kket{\downarrow}_j)/\sqrt{2}$.
In this orthonormal basis, the averaged Haar random gate takes the form
\begin{align}
    \mathcal{U}_{\text{Haar}} = \mathcal{P} = \bigotimes_{j = 1}^L \sum_{\{\sigma_j = \uparrow,\downarrow\}}\kket{\sigma_j}\bbra{\sigma_j}_j,
\end{align}
which is exactly the projector onto the effective Hilbert space.

Second, the averaged long-range unitary gate $\mathcal{U}_{\text{LR}}$ in $\mathcal{H}^{(2)}$ can be derived by Gaussian integration
\begin{align}
    \mathcal{U}_{\text{LR}} = \overline{U_{\text{LR}} \otimes U_{\text{LR}}^* \otimes U_{\text{LR}} \otimes U_{\text{LR}}^*} = \exp\left(-\sum_{1\leq i < j \leq L}\frac{K}{2|i-j|^{\alpha}}\left(\sum_{\mu = 1,2} Z_{i,\mu}Z_{j,\mu} - Z_{i,\bar{\mu}}Z_{j,\bar{\mu}}\right)^2\right).
\end{align}
Finally, averaged measurement $\mathcal{M}$ takes the form
\begin{align}
    \mathcal{M} = (1-p)\, \mathds{1}^{\otimes 4} + p \sum_{m = \pm} P_{j,m}^{\otimes 4},
\end{align}
where $P_{j,\pm} = (1 \pm Z_j)/2$ are the projectors associated with measurements in the computational basis.

To formulate an exact mapping, we work in a continuous time limit with $p = \Gamma \delta t$, $K = J\delta t$ and $\delta t \to 0$~\cite{bao2021symmetry}.
In this limit, $\mathcal{U}_{\text{LR}}$ and $\mathcal{M}$ deviates from the identity operator by $O(\delta t)$, and we can further project $\mathcal{U}_{\text{LR}}$ and $\mathcal{M}$ onto the effective Hilbert space
\begin{align}
    \mathcal{P}\mathcal{U}_{\text{LR}}\mathcal{M}\mathcal{P} = \mathcal{P} \mathcal{U}_{\text{LR}}\mathcal{P} \cdot \mathcal{P} \mathcal{M}\mathcal{P} + O(\delta t^2),
\end{align}
where $\mathcal{P}\mathcal{U}_{\text{LR}}\mathcal{P} = e^{-\delta t H_{\text{eff}}(\mathcal{U}_{\text{LR}}) + O(\delta t^2)}$ and $\mathcal{P}\mathcal{M}\mathcal{P} = e^{-\delta t H_{\text{eff}}(\mathcal{M})+ O(\delta t^2)}$.
Thus, remarkably, the hybrid dynamics in each time step generates an imaginary time evolution in the effective Hilbert space:
\begin{align}
    \mathcal{P}\kket{\tilde{\rho}(t+\delta t)} = e^{-\delta tH_{\text{eff}} + O(\delta t^2)} \mathcal{P}\kket{\tilde{\rho}(t)},
\end{align}
where $H_{\text{eff}} = H_{\text{eff}}(\mathcal{U}_{\text{LR}}) + H_{\text{eff}}(\mathcal{M})$.
In the long time limit $t = N_T\delta t \to \infty$ of the circuit, the replicated density matrix evolves to the ground state of $H_{\text{eff}}$, i.e. $\lim_{t \to \infty} \kket{\tilde{\rho}(t)} = \lim_{t \to \infty} e^{-tH_{\text{eff}}}\kket{\rho_0} = \kket{\psi_{gs}}$.
Plugging in the above expressions into Eq.~\ref{eq:evolution}, we can explicitly determine the effective Hamiltonian
\begin{align}
    H_{\text{eff}} = \sum_{i < j} -\frac{J}{|i - j|^\alpha}(3 \sigma^z_i \sigma^z_j - \sigma^x_i \sigma^x_j) - \sum_j h \sigma^x_j,
\end{align}
where $h = \Gamma/3 + \sum_{r = 1}^{\infty} J/(9r^\alpha)$.
This is simply a long-range Ising model, with a symmetry breaking phase-transition described by a long wavelength theory with the action~\cite{fisher1972critical}
\begin{align}
    \mathcal{S} = \int \frac{\mathrm{d} q\mathrm{d}\omega}{(2\pi)^2} \frac{1}{2}\left(g\omega^2 + r + aq^{\alpha-1} + bq^2\right)\phi_{\mathbf{q}}\phi_{\mathbf{-q}} + u\int\prod_{i = 1}^3\frac{\rd q_i\rd \omega_i}{(2\pi)^2}\phi_{\mathbf{q}_1}\phi_{\mathbf{q}_2}\phi_{\mathbf{q}_3}\phi_{-\mathbf{q_1}-\mathbf{q_2}-\mathbf{q_3}},
\end{align}
where $\mathbf{q} = (q,\mathrm{i}\omega)$ with $\omega$ denoting the Matsubara frequency.
As discussed in the manuscript, this field theory shows long-range interactions become relevant for $\alpha=3-\eta$, where $\eta$ is the short-range anomalous dimension.
We can additionally obtain an approximation for $z(\alpha)$ by demanding $\int \mathrm{d} q\mathrm{d}\omega \omega^2 \phi_{\mathbf{q}}\phi_{\mathbf{-q}}$ and $\int \mathrm{d} q\mathrm{d}\omega q^{\alpha-1}\phi_{\mathbf{q}}\phi_{\mathbf{-q}}$ have the same scaling dimension for $\alpha < 3-\eta$, yielding $z=(\alpha-1)/2$.
Thus, the field theory additionally predicts $z(\alpha)$ decreases with $\alpha < 3-\eta$, as observed in our numerics (Fig.~2).

Finally, we demonstrate precisely how the phase transition in the ground state of $H_{\text{eff}}$ relates to the subsystem entanglement entropy $S^{(2)}_A$ in Eq.~\eqref{eq:nth_entropy_eff}. 
In the effective Hilbert space, the swap operator becomes $\mathcal{C}_{\ell,A} = \prod_{i\in A}\sigma^x_i$, which inserts a pair of domain walls at the edges of subsystem $A$.
Therefore, $S^{(2)}_A$ can be expressed as
\begin{align}
    e^{-S_A^{(2)}} = \frac{\bbrakket{\mathcal{I}}{\prod_{i\in A} \sigma^x_i |\psi_{gs}}}{\bbrakket{\mathcal{I}}{\psi_{gs}}}.
    \label{eq:2nd_entropy_eff}
\end{align}
The reference state $\kket{\mathcal{I}}$ is a symmetry breaking product state given by $\kket{\mathcal{I}} = \bigotimes_{i = 1}^L [(\sqrt{3}+1)\kket{\uparrow} + (\sqrt{3}-1)\kket{\downarrow}]/\sqrt{2}$.
As established in Ref.~\cite{bao_theory_2020,jian2020measurement,bao2021symmetry}, the $\mathbb{Z}_2$ Ising symmetry is spontaneously broken in the entanglement phase transition.
In the area-law phase, the ground state preserves the Ising symmetry and is a condensate of the domain-wall operators.
Therefore, the matrix element on the right-hand side of Eq.~\eqref{eq:2nd_entropy_eff} is a constant independent of the subsystem size $|A|$.
In the volume-law phase,  due to the domain-wall insertion at the edges of $A$, the matrix element decays exponentially in $|A|$, indicating a volume-law scaling entanglement entropy.
The change in the universality of the phase transition in the ground state thus manifests in the matrix element $e^{-S_A^{(2)}}$, resulting in a change in the universality of the entanglement transition.

\section{Finite Size Scaling Summary} \label{sec:fss-summary}

\begin{figure}[h!]
	\centering
	\includegraphics[width=1.0\linewidth]{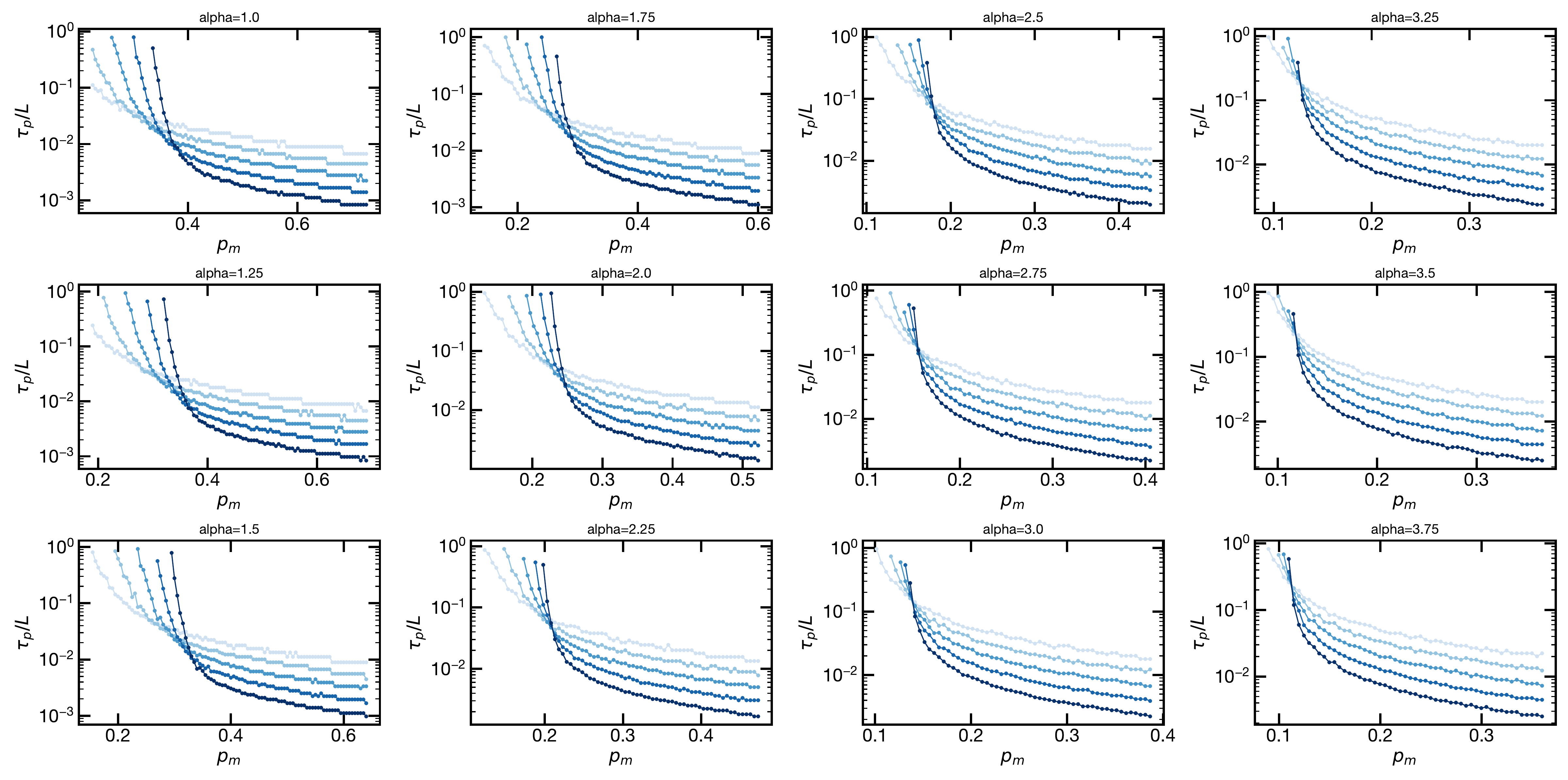}
	\caption{Un-collapsed $\tau_p(p)$ data. Light blue indicates small system sizes, and darker blue indicates larger system sizes, ranging from $L=32,\cdots512$.}
	\label{fig:ptime-raw}
\end{figure}

\begin{figure}[h!]
	\centering
	\includegraphics[width=1.0\linewidth]{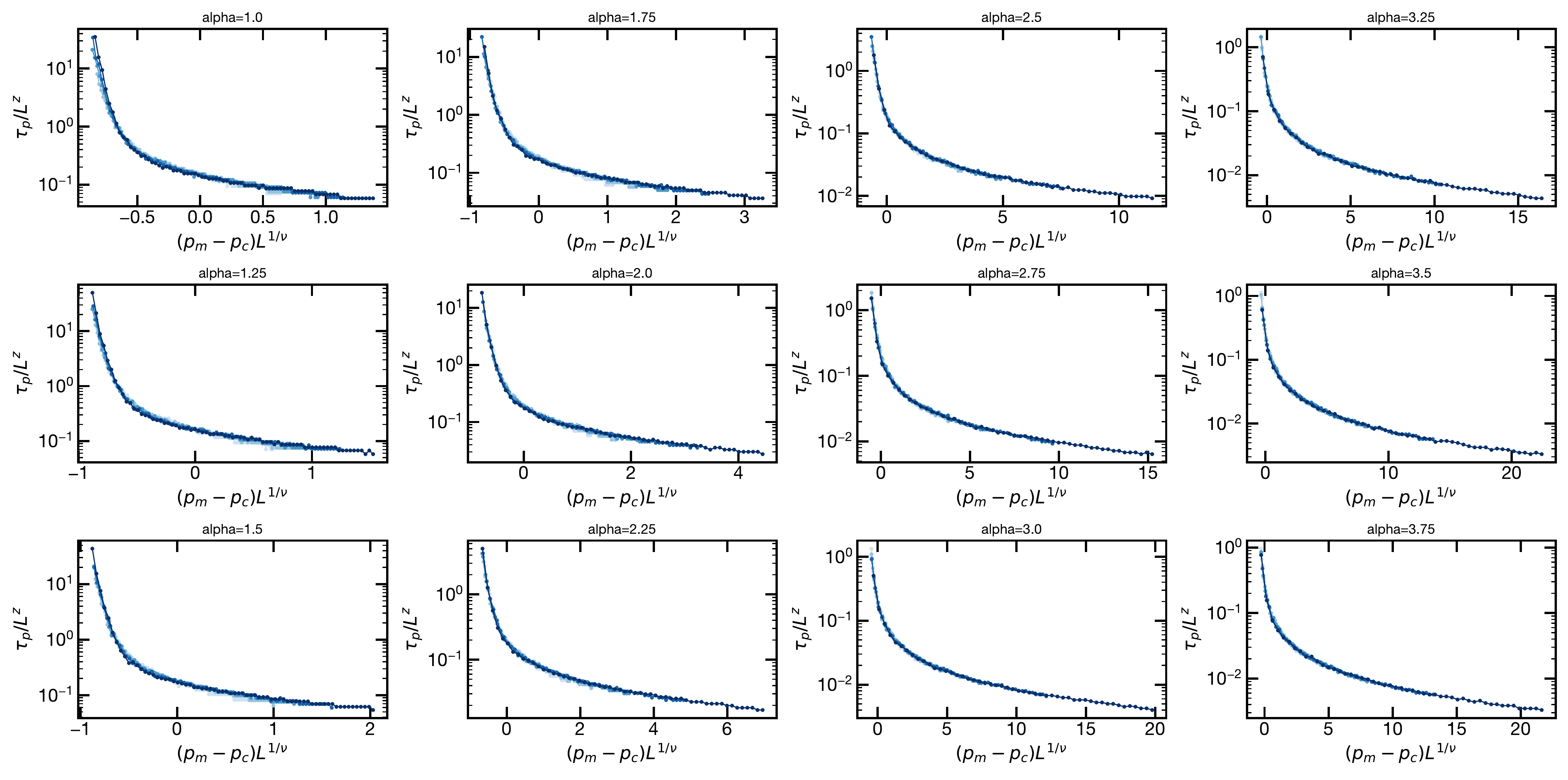}
	\caption{Collapsed $\tau_p(p)$ data. Light blue indicates small system sizes, and darker blue indicates larger system sizes, ranging from $L=32,\cdots512$. The quality of the collapse is consistent across $\alpha$.}
	\label{fig:ptime-collapse}
\end{figure}

\begin{figure}[h!]
	\centering
	\includegraphics[width=1.0\linewidth]{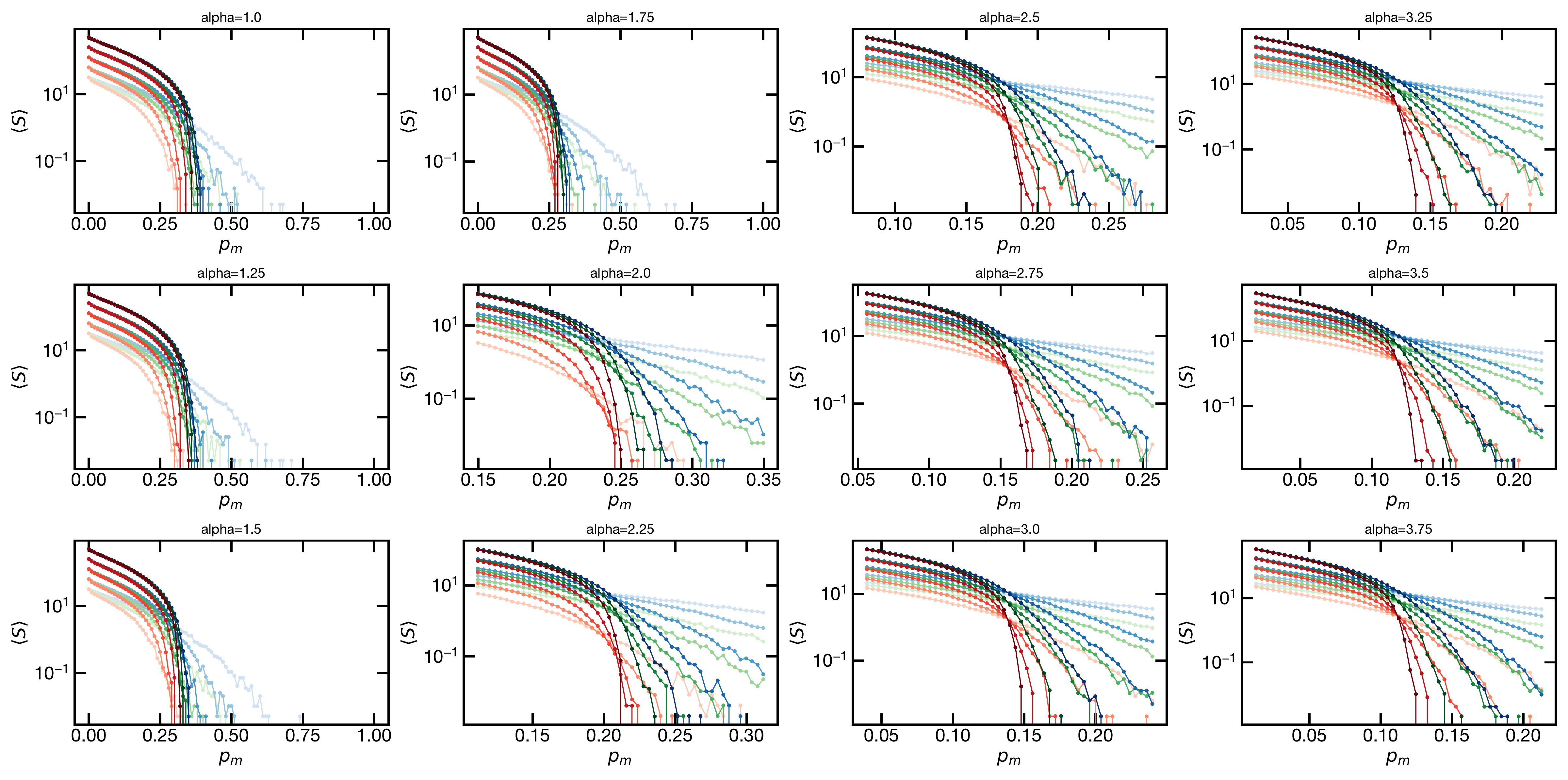}
	\caption{Un-collapsed $S(p, t)$. As in the main text, different colors indicate different circuit depths, with \{blue, green, red\} $\leftrightarrow T=\{1/2, 2/3, 2\} \cdot L$.}
	\label{fig:global-s-raw}
\end{figure}

\begin{figure}[h!]
	\centering
	\includegraphics[width=1.0\linewidth]{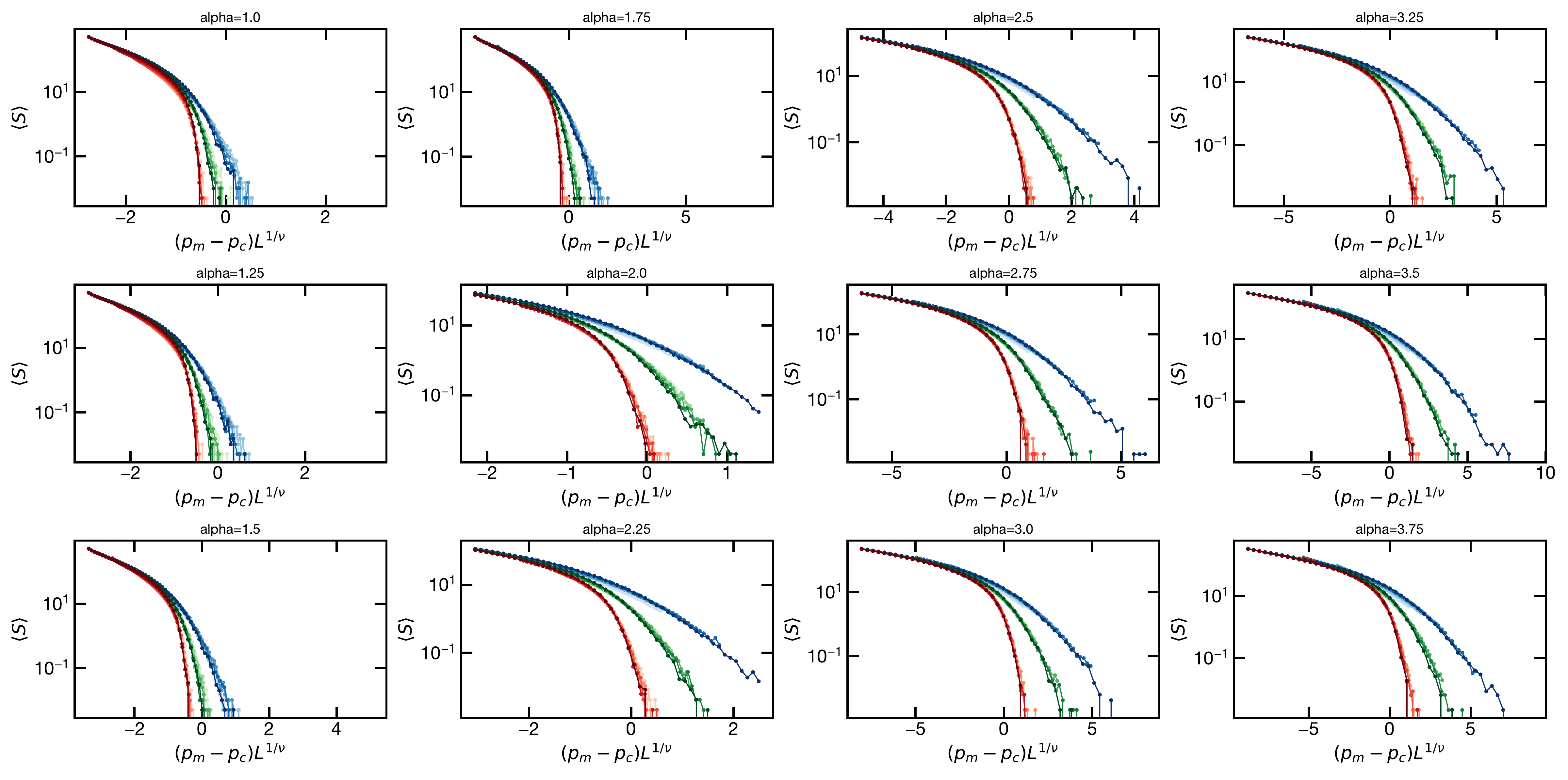}
	\caption{Collapsed $S(p,t)$, using the critical exponents extracted from $\tau_p$. As in the main text, different colors indicate different circuit depths, rescaled according to $z$, with \{blue, green, red\} $\leftrightarrow T=\{1/2, 2/3, 2\} \cdot L^z$.}
	\label{fig:global-s-collapse}
\end{figure}

\begin{figure}[h!]
	\centering
	\includegraphics[width=0.7\linewidth]{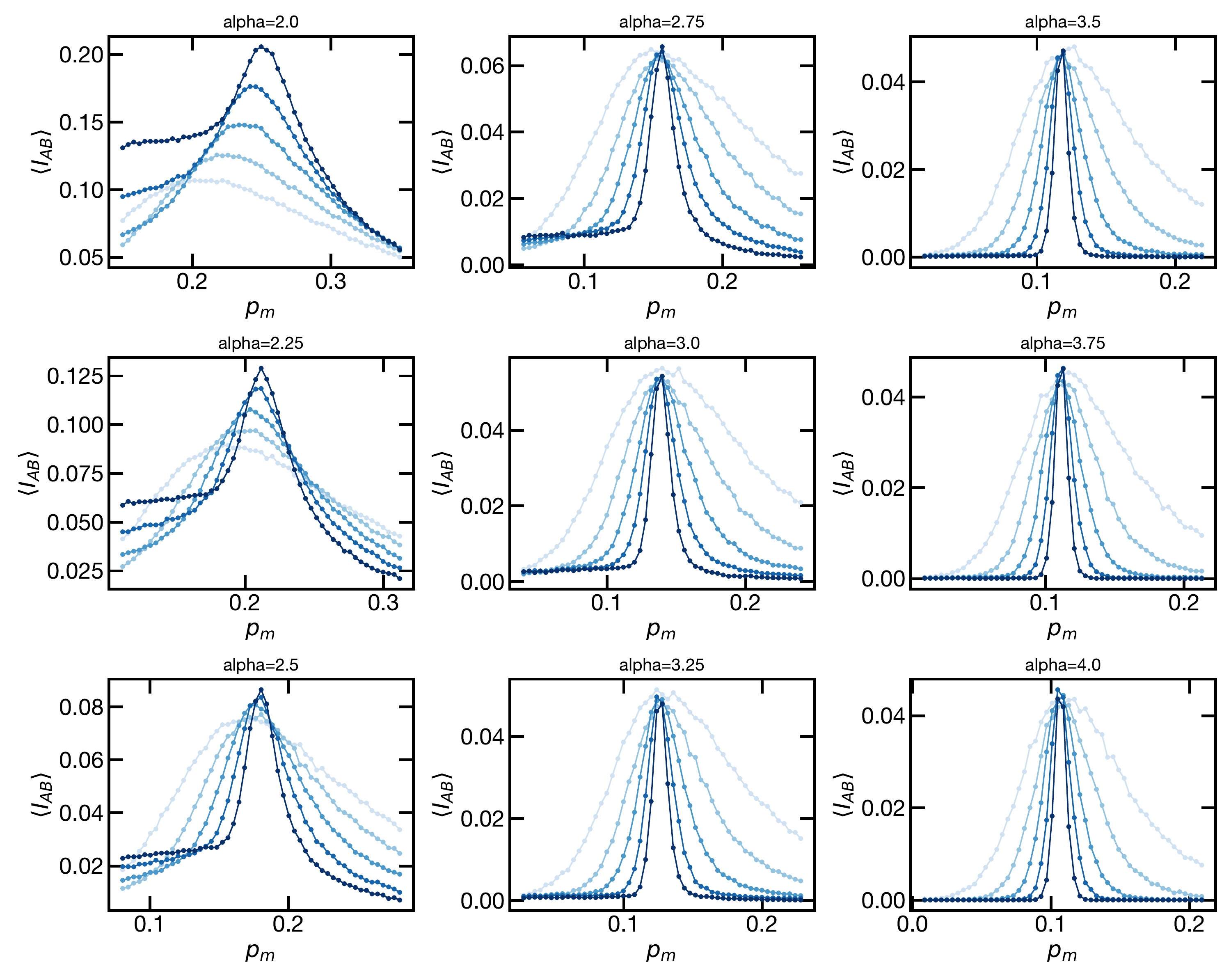}
	\caption{Un-collapsed $I_{AB}(p)$ data. Light blue indicates small system sizes, and darker blue indicates larger system sizes, ranging from $L=32,\cdots512$. The scaling of the $I_{AB}$ peak with system size can be observed for $\alpha < 3$.}
	\label{fig:ap-mi-raw}
\end{figure}

\begin{figure}[h!]
	\centering
	\includegraphics[width=0.7\linewidth]{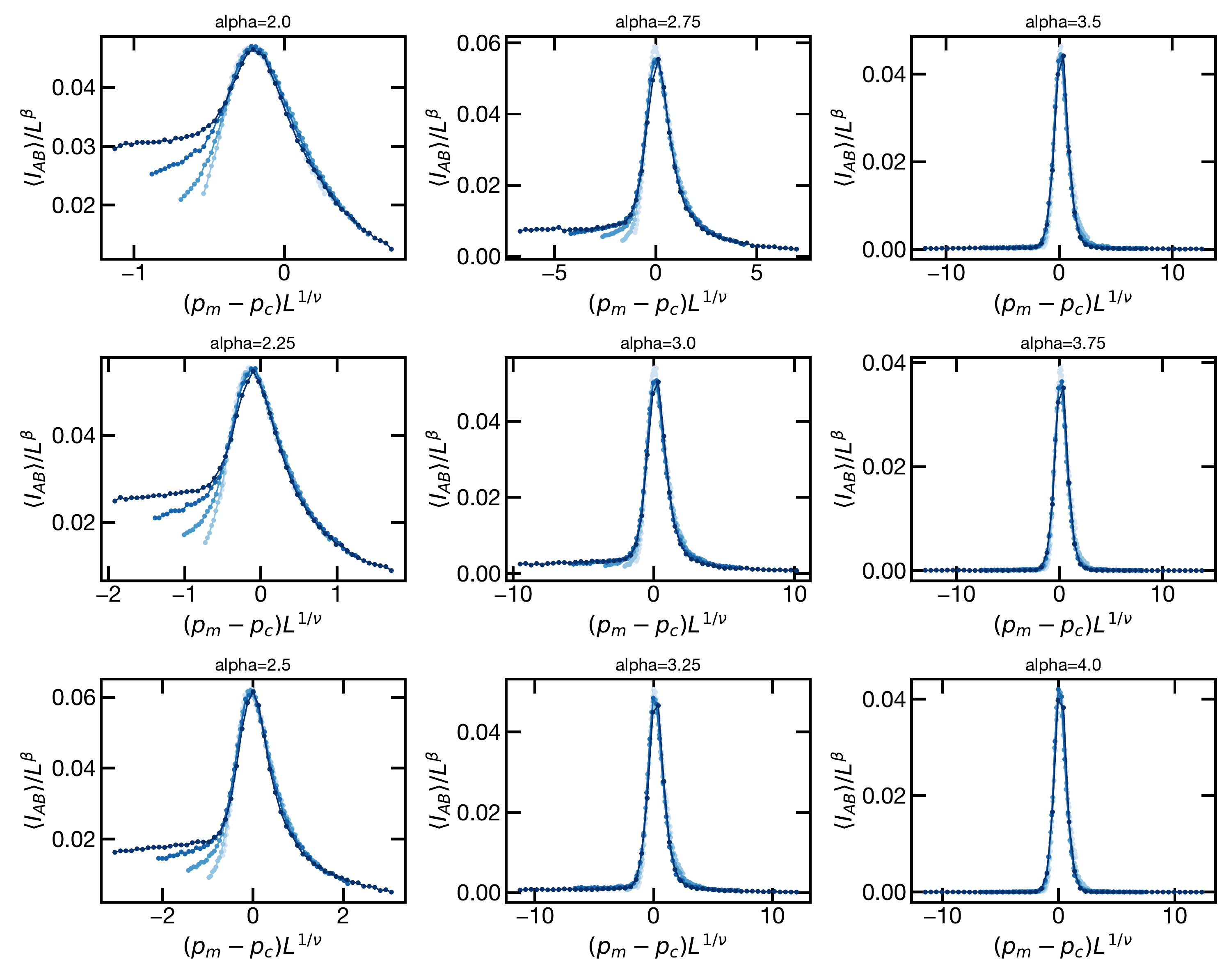}
	\caption{Collapsed $I_{AB}(p)$ data. Light blue indicates small system sizes, and darker blue indicates larger system sizes, ranging from $L=32,\cdots512$.}
	\label{fig:ap-mi-collapse}
\end{figure}

\bibliographystyle{apsrev4-2}
\bibliography{bibliography}